\documentclass[sigconf]{acmart}

\AtBeginDocument{%
  }

\setcopyright{acmcopyright}
\copyrightyear{2018}
\acmYear{2018}
\acmDOI{XXXXXXX.XXXXXXX}

\acmConference[ACM MM'23]{ACM Multimedia}{Oct 28--Nov 05,
  2023}{Ottawa, Canada}

\acmPrice{15.00}
\acmISBN{978-1-4503-XXXX-X/18/06}

\acmSubmissionID{409}

\usepackage{graphicx}
\usepackage{amsmath}
\usepackage{booktabs}
\usepackage{multirow}
\usepackage{float}
\usepackage{adjustbox}

\copyrightyear{2023} 
\acmYear{2023} 
\setcopyright{acmlicensed}\acmConference[MM '23]{Proceedings of the 31st ACM International Conference on Multimedia}{October 29-November 3, 2023}{Ottawa, ON, Canada}
\acmBooktitle{Proceedings of the 31st ACM International Conference on Multimedia (MM '23), October 29-November 3, 2023, Ottawa, ON, Canada}
\acmPrice{15.00}
\acmDOI{10.1145/3581783.3611787}
\acmISBN{979-8-4007-0108-5/23/10}

\begin{document}

\title{Towards Accurate Lip-to-Speech Synthesis in-the-Wild}


\author{Sindhu Hegde}
\authornote{Both authors have contributed equally to this research.}
\authornote{This work was done while pursuing MS by Research at IIIT Hyderabad.}
\email{sindhu@robots.ox.ac.uk}
\affiliation{%
\institution{VGG, University of Oxford}
\country{United Kingdom}
}

\author{Rudrabha Mukhopadhyay}
\authornotemark[1]
\email{radrabha.m@research.iiit.ac.in}
\affiliation{
\institution{CVIT, IIIT Hyderabad}
\country{India}
}

\author{C.V Jawahar}
\email{jawahar@iiit.ac.in}
\affiliation{%
\institution{CVIT, IIIT Hyderabad}
\country{India}
}

\author{Vinay Namboodiri}
\email{vpn22@bath.ac.uk}
\affiliation{%
\institution{University of Bath}
\country{United Kingdom}
}

\renewcommand{\shortauthors}{Sindhu Hegde, Rudrabha Mukhopadhyay, C.V Jawahar, \& Vinay Namboodiri}

\begin{abstract}
In this paper, we introduce a novel approach to address the task of synthesizing speech from silent videos of any in-the-wild speaker solely based on lip movements. The traditional approach of directly generating speech from lip videos faces the challenge of not being able to learn a robust language model from speech alone, resulting in unsatisfactory outcomes. To overcome this issue, we propose incorporating noisy text supervision using a state-of-the-art lip-to-text network that instills language information into our model. The noisy text is generated using a pre-trained lip-to-text model, enabling our approach to work without text annotations during inference. We design a visual text-to-speech network that utilizes the visual stream to generate accurate speech, which is in-sync with the silent input video. We perform extensive experiments and ablation studies, demonstrating our approach's superiority over the current state-of-the-art methods on various benchmark datasets. Further, we demonstrate an essential practical application of our method in assistive technology by generating speech for an ALS patient who has lost the voice but can make mouth movements. Our demo video, code, and additional details can be found at \url{http://cvit.iiit.ac.in/research/projects/cvit-projects/ms-l2s-itw}.
\end{abstract}


\begin{teaserfigure}
\centering
  \includegraphics[width=0.9\textwidth]{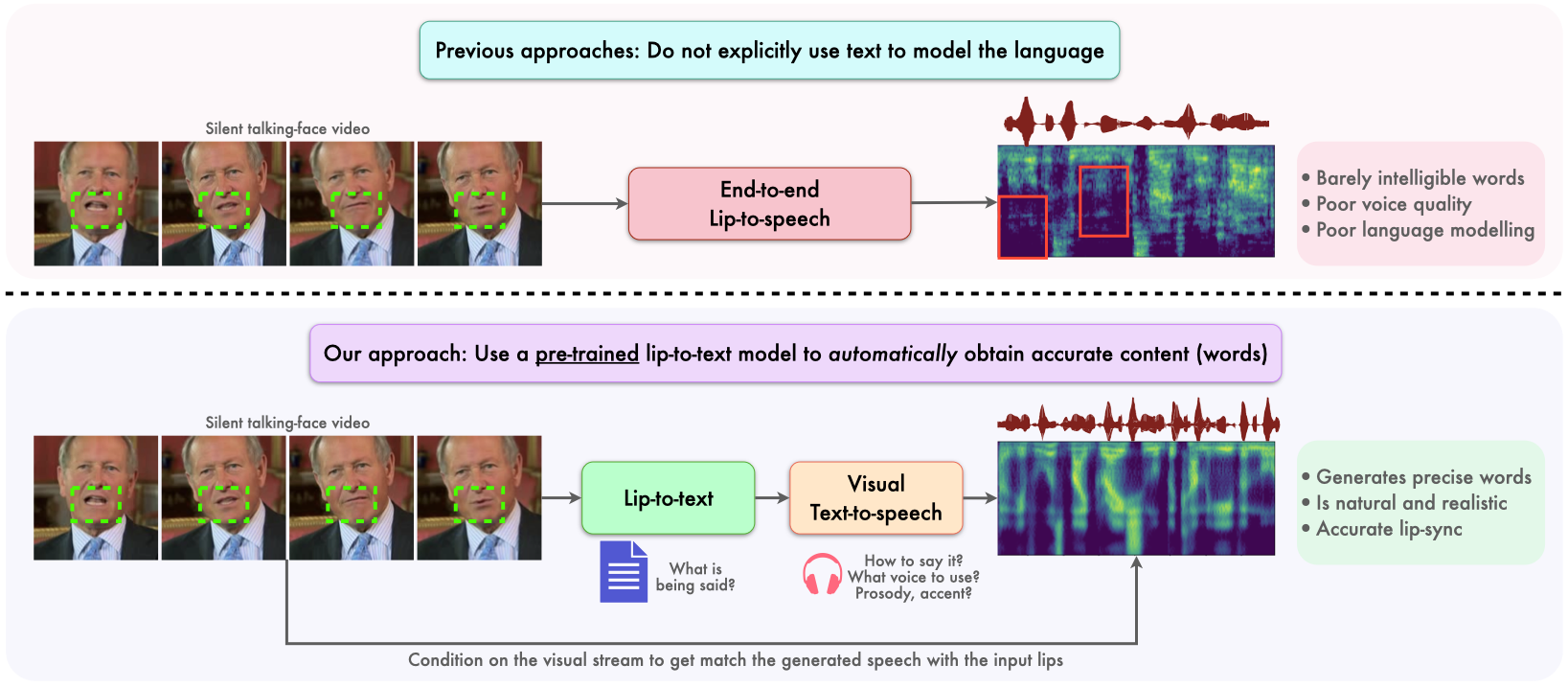}
  \caption{We propose a novel approach for multi-speaker lip-to-speech synthesis in the wild. Prior works try to learn a language model directly from raw speech, which only provides weak supervision due to the presence of other acoustic variations such as voice, accents, and prosody. We solve this problem by relying on recent advancements in lip-to-text generation. We condition on the noisy text outputs and lip video to generate natural speech with clearly pronounced words.}
  \Description{}
  \label{fig:teaser}
\end{teaserfigure}


\maketitle

\section{Introduction}
\label{sec:intro}


As multi-sensory beings, humans have the ability to experience the world through various senses, adapting to make the most of what we perceive. When one sense is diminished, like hearing, the brain compensates by relying more on other senses, like vision, to maintain independence and navigate the environment. One of the most prevalent ways for deaf individuals to comprehend speech is through lip reading - the process of understanding spoken content purely from silent lip movements. Therefore, lip reading has numerous applications, ranging from searching through old silent films to being used as an assistive technology for people who are unable to speak. For example, people with Amyotrophic Lateral Sclerosis (ALS) often lose their ability to speak due to issues with their vocal cords but can mouth words. Therefore, developing automated lip reading technologies can improve the communication abilities of people with ALS. In the past few years, previous works have shown that it is far easier for deep learning models to perform lip reading than it is for humans~\cite{chung2017lip}. There has been a growing interest in the research community~\cite{Afouras2018DeepLR, afouras2018lrs3, chung2016lip, chung2017lip} to solve the task of transcribing silent lip movements into text. While giant strides are being made to solve lip-to-text generation, only a handful of works go a step further to generate speech from silent lip movements. The widening gap between these two closely related tasks is becoming increasingly evident. In other words, while lip-to-text models are reaching word error rates as low as $17\%$ WER~\cite{serdyuk22_interspeech}, there are no practically usable lip-to-speech systems that generate natural, meaningful speech for in-the-wild identities. 

\paragraph{Applications for lip-to-speech synthesis}
Lip-to-Speech generation is not only more exciting and challenging but also has more far-reaching applications. For instance, having a live video call with a person who has lost their voice is more engaging than reading the text transcription of their silent lip movements. Speech can convey more: it is possible to express emotions through silent lip movements if it is converted to speech rather than text. Speech is also more instantaneous for the listener: it is easier and faster to hear than to read text. The immersive user experience is a significant impetus for developing algorithms that translates lip movements directly to speech. Lip-to-Speech can also be applied in forensic investigations, where silent footage can be analyzed to determine what was being said or generate speech from archival film footage. Our work showcases the practical impact of lip-to-speech generation by generating speech for the silent lip movements of an ALS patient, a feat not previously demonstrated by other models in this field. We also show results on reading mouth movements of people suffering from hearing loss to demonstrate the effectiveness of accurate lip-to-speech synthesis further. More details about the applications can be found in Section~\ref{sec:applications_l2s}.

\paragraph{The challenges in lip-to-speech synthesis}
However, synthesizing speech from silent talking-face videos is far more challenging than lip-to-text generation. Not only does the model first needs to distill the content from the lip movements, but it also needs to generate the final speech by accurately modeling the speaker attributes like voice, accent, and style. Owing to these difficulties, until recently, most of the existing works~\cite{ephrat2017vid2speech, vougioukas2019video, Prajwal_2020_CVPR} focused on building either single-speaker models or constrained multi-speaker models with limited corpus size as well as limited vocabulary. While more recent efforts like~\cite{Prajwal_2020_CVPR} have extended lip-to-speech to ``in-the-wild" environments, most of these models are speaker-specific in nature, i.e., they only work on speakers they are trained on. The speaker-independent models~\cite{msl2s, vcagan, svts, multitask-l2s} suffer from numerous weaknesses and fail to accurately learn language and speech attributes like voice, prosody, etc. Due to the sheer difficulty of the task, they turn out to be well below expectations for an end-user application.

\subsection{Our Contributions}
First, we would like to clearly state that our goal is to be able to generate speech for a silent lip video of \textit{any} speaker in the wild and in \textit{any} desired target voice. The fundamental premise of this paper is that current models trying to solve this task struggle to learn language and speech attributes because they try to learn both of these solely from speech supervision. They inadvertently disregard the advancements that have been made in the sibling task of the lip-to-text generation. We argue that learning lip-to-speech without any understanding/supervision of text is very challenging to achieve, which can be observed in all the current models, where they fail to produce clear and intelligible outputs in unconstrained settings. Our key idea is to use the fact that lip-to-text models already have learned the task of extracting content from lips, as they are trained with text supervision. We show that we can achieve accurate, natural, and high-quality speech outputs by using the silent video input and the noisy text transcriptions from a pre-trained lip-to-text model. We train a visual text-to-speech model to synthesize speech that syncs with the input video. Our approach outperforms the existing multi-speaker lip-to-speech models by a significant margin in terms of both qualitative and quantitative evaluations. We highlight our key contributions below. 

\begin{enumerate}
\item Lip-to-Speech generation is challenging with only speech supervision, and current models struggle to learn language and speech attributes. To address this, we propose using a pre-trained lip-to-text model's output to aid in lip-to-speech generation.

\item Our approach involves training a visual text-to-speech model to synthesize speech that syncs with the input video. Using this approach, we outperform existing multi-speaker lip-to-speech models in qualitative and quantitative evaluations.

\item Our proposed approach enables accurate and natural speech output for silent lip videos of any speaker in any desired target voice. 
\end{enumerate}

\section{Related Works}

\paragraph{Lip-to-Speech}
Generating speech directly from lip movements has always been a challenge for researchers due to the ill-posed nature of the task. While recognizing content from lip movements itself is challenging due to the one-to-many relation present between visemes and phonemes, lip-to-speech synthesis also needs to model speech-related variations like voice, accents, and prosody. Therefore, it was initially attempted in laboratory setups~\cite{cooke2006audio} using datasets with a minimal vocabulary. The initial works to~\cite{Ephrat2017ImprovedSR, ephrat2017vid2speech, vougioukas2019video, Akbari2017Lip2AudspecSR} used fully convolutional neural networks to learn a mapping between lip movements and speech. These models were also trained on speaker-specific data and did not work for unseen speakers. Moreover, such models were incapable of handling ``in-the-wild" settings. To mitigate this issue, a sequence-to-sequence model, Lip2Wav~\cite{Prajwal_2020_CVPR}, was proposed that was trained on large amounts of speaker-specific data taken from in-the-wild YouTube videos. Lip2Wav worked reasonably well on speakers seen during training but did not extend to handling unseen speakers. Several more recent studies~\cite{msl2s, vcagan, multitask-l2s, svts} have attempted to develop end-to-end lip-to-speech synthesis models on large datasets containing data from hundreds of speakers. For instance, in~\cite{msl2s}, authors proposed a variational approach that matches the distributions of lip movements and speech segments to project them into a shared space, which allows for handling the high variations of in-the-wild speakers to some extent. Meanwhile, both~\cite{vcagan, svts} utilized a transformer-based approach to convert lip-to-speech synthesis into a sequence-to-sequence problem, where a sequence of lip movements is translated into a sequence of speech tokens. Additionally,~\cite{multitask-l2s} treated lip-to-speech synthesis as a multi-tasking problem, where the model predicts text transcripts in addition to speech forcing the model to learn better content information. However, all of these models produce sub-par outputs with words that are barely uttered and often unintelligible with unnatural voice and prosody. We argue that these works are limited because they do not use the recent advancements in the sibling task, lip-to-text, which we discuss below.

\paragraph{Lip-to-Text}
Lip reading has become synonymous with lip-to-text, and several advancements have been made to achieve highly accurate text transcriptions of silent lip videos. The very first works~\cite{LipNet_arxiv_2016} used datasets~\cite{cooke2006audio, harte2015tcd} collected in a laboratory setup and trained simple RNN-based architectures. One of the initial approaches~\cite {chung2016lip} handling the ``in-the-wild" setup posed the problem in a word-level classification setting and continued using RNN-based models. Subsequent approaches~\cite{chung2017lip} have attempted to tackle this task by posing it as a sequence-to-sequence problem. The goal here is to translate a sequence of lip movements into a sequence of characters. Approaches like~\cite{Afouras2018DeepLR} compared multiple sequence-to-sequence architectures, including transformers, to achieve favorable word error rates (WER). A more recent work~\cite{Prajwal_2022_CVPR} uses a variant of transformer and specific attention to the mouth region through visual transformer pooling and achieves state-of-the-art results in lip reading. Other works such as AV-HuBERT~\cite{av-hubert} have also improved the state-of-the-art for in-the-wild lip reading, paving the way for practical applications. Our work demonstrates the importance of pre-trained lip-to-text models in achieving precise lip-to-speech synthesis in uncontrolled environments.

\paragraph{Text-to-speech}
Our proposed approach uses text as an intermediate output and converts it into speech. Text-to-Speech has long interested the speech community and has historically seen a lot of progress. Modern deep learning architectures~\cite{tacotron2, deepvoice3, fastspeech2} enable several industrial and multimedia applications. They are accurate in terms of uttered content and have natural speaking styles, prosody, and voice. The efforts slowly shifted towards multi-speaker scenario, which is far more challenging due to the sheer variation in prosody, voice, and other speaker-related attributes like pitch and tone. Most of these methods are provided with a separate voice token~\cite{sv2tts} containing identity information about the target speaker. The voice tokens are generated from a short speech segment of a target speaker and often only capture the voice information while failing to capture the prosody. Thus, there has also been a push for using additional information like lip movements~\cite{vdtts, neural_dubber} to help TTS models generate higher-quality outputs. Since lip-to-speech synthesis requires us to generate accurate prosody and speaking style, our approach is inspired by this line of work. We design a visual text-to-speech model that works uses both text and visual features from state-of-the-art lip reading network~\cite{Prajwal_2020_CVPR} to generate natural, accurate speech. Our Visual TTS is integral to generating speech that perfectly syncs with the input silent lip video.

\section{Lip-to-Speech Synthesis in the Wild}

\begin{figure*}[t]
  \includegraphics[width=0.9\textwidth]{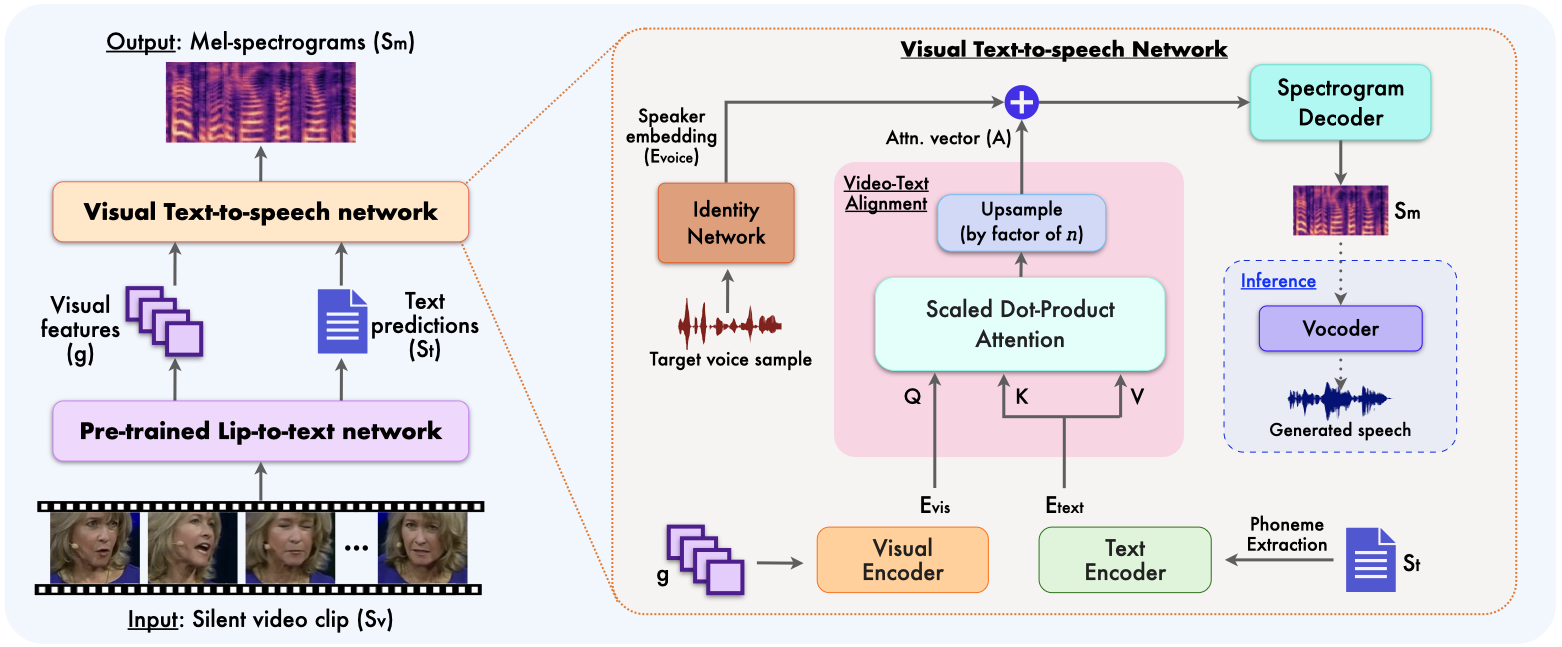}
  \vspace{-10pt}
  \caption{Overview of our approach. We first extract the visual features and the text predictions from a pre-trained lip-to-text network. Using a visual text-to-speech (TTS) model, we can generate speech outputs that sync with the silent video input. The visual TTS encodes the visual and textual (in the form of phonemes) inputs and aligns them in time using the scaled dot-product attention. For each query video time-step, we retrieve the phoneme to utter at that time using this attention mechanism. After adding the speaker identity embedding, these are then upsampled and decoded into melspectrograms. The melspectrograms are converted into natural waveforms using a pre-trained vocoder.}
  \label{fig:architecture}
  \vspace{-10pt}
\end{figure*}

We start the discussion by highlighting the key issues and challenges in the existing lip-to-speech works, leading to their poor performance in unconstrained settings. We then present our framework with a detailed description of the modules involved.

\subsection{Issues and challenges in existing works}

\subsubsection{Learning language from speech}
As discussed previously, all the current lip-to-speech works directly generate speech from the lips. It is known that learning a language model is crucial for reading the lips accurately. However, the current multi-speaker lip-to-speech models are sub-par because they try to learn a language model in the speech modality, which contains a large diversity of speaker identities, styles, accents, and prosody. Thus, we argue that we need some other way of incorporating language knowledge. 

\subsubsection{The missing block of lip-to-speech: lip-to-text}
Lip-to-Speech synthesis models have two tasks: (i) inferring the content from lips and (ii) inferring the style in which that content is spoken. If we have the content that is being spoken, then our task is now reduced to just generating the speech that matches the silent lip video. This is the premise of our paper. But how do we get this text information, especially when we only have a silent lip video as input? We show that we can get this text information from pre-trained lip-to-text models: a class of models closely related to our task at hand but have been mainly ignored in previous works on lip-to-speech synthesis. We design an approach that can build upon the current works~\cite{Afouras2018DeepLR, Prajwal_2022_CVPR} in lip reading, i.e., pre-trained lip-to-text models, and generate far more accurate speech outputs. 

\subsubsection{Achieving accurate lip-sync}
\label{sec:lipsync}
Now that we have the text, the next step is to generate the ``right" kind of speech output. That is, the generated speech must match the input lip sequence. There are many ways to utter the same sentence, but only one will match the input lip video. Thus, it is worth noting that a trivial text-to-speech will not serve our task. Instead, we need a text-to-speech model that is also conditioned on video input. 

\subsection{Our Approach}
The task of lip-to-speech synthesis is formulated as follows: given a silent video frame sequence of a person talking $S_v = \{I_1, I_2, ..., I_p\}$, the goal is to generate the corresponding speech melspectrogram sequence, $S_m = \{M_1, M_2, ..., M_q\}$. An overview of our proposed two-stage framework is depicted in Figure~\ref{fig:architecture}. We use a state-of-the-art lip-to-text model~\cite{Prajwal_2022_CVPR} to obtain the lip features and noisy text transcriptions for the given silent lip video $S_v$. We design a visual text-to-speech model conditioned on (i) the noisy text and (ii) the lip features to produce high-quality speech outputs $S_m$ that are in sync with the input silent lip video. This solves the task of lip-to-speech synthesis.

\subsubsection{Pre-trained Lip-to-Text}
Lip-to-Text networks ingest the silent talking face videos as input and transcribe what is being uttered. The lip-to-text models~\cite{DeepAVLR_tpami_2018, chung2016lip, chung2017lip, afouras2018lrs3, Afouras2018DeepLR, Prajwal_2022_CVPR} are now capable of generating text transcripts with word error rates as low as $20-30\%$. One such recent work trained on public datasets is VTP~\cite{Prajwal_2022_CVPR}. Specifically, this work stands out due to its two unique aspects: (i) it has been shown to be data efficient, (ii) it contains a strong visual backbone that can attend to and extract accurate lip features. These visual features have been shown to work well in several other tasks such as visual keyword spotting~\cite{KR2021VisualKS} and even spotting mouth movements in sign language~\cite{sign_language}. The sub-word units considered to learn the text representations help to model the ambiguities of the task significantly better than using characters. Also, the sub-word tokens are semantically meaningful and provide a language prior, thereby resulting in a significant performance boost. In addition, the visual representations learned using the backbone network track and aggregate the spatio-temporal lip movement features, mainly due to the strong attention-based pooling mechanism. We thus propose to adopt this model as our first module to generate text and visual features from silent lip videos.  

\paragraph{Overview of VTP~\cite{Prajwal_2022_CVPR}:} The model inputs a sequence of $5$ consecutive frames, $S_v \in \mathbb{R}^{T \times H \times W \times 3}$ ($T = 5$) and extracts low-level features using a spatio-temporal residual CNN block. These individual frame-wise features are then processed using the visual transformer pooling block, which consists of a series of transformer layers. The output from this block, $z_t$ is a self-attended feature map, which is used along with a learnable query vector, $Q_{att}$ to obtain the spatially weighted average of $z_t$. These compact per-frame visual representations are stacked along the temporal dimension to obtain a temporal embedding sequence, $g \in \mathbb{R}^{T \times fd}$, where $fd$ is the transformer feature dimension. Finally, the learned visual representations are passed to the transformer encoder-decoder network to predict the text outputs in the form of sub-word tokens in an auto-regressive fashion. The model uses beam search decoding~\cite{beam} and language model rescoring~\cite{las} strategies to obtain the final sentence outputs $S_t$. We refer the reader to~\cite{Prajwal_2022_CVPR} for more details on the architecture and training strategies.

\vspace{-5pt}
\paragraph{Adopting VTP for our task:} For our task at hand, we use the pre-trained VTP network to obtain: (i) text predictions - the final decoded output of the model $S_t$, and (ii) per-frame visual representations - $g$. The text predictions directly act as input to our speech generation module. The visual representations serve as a condition for speech generation, which is crucial to obtain speech that is in-sync with the silent input video. 

\subsubsection{Visual Text-to-Speech}
Once we have the accurate text predictions, the next step is to generate the corresponding speech sequence $S_m$, which is in-sync with the input video clip $S_v$. As explained in Section~\ref{sec:lipsync}, directly using state-of-the-art TTS models to synthesize speech from text inputs will generate out-of-sync speech that does not match the input video. We thus design a TTS network by conditioning the model on the input video features. Our visual TTS network majorly comprises five components: (i) Text Encoder, (ii) Visual Encoder, (iii) Visual-Text Attention, (iv) Speaker Embedding, and (v) Spectrogram Decoder. We delve into each of these components below.

\paragraph{Text Encoder:}
As followed in most of the TTS networks~\cite{fastspeech2}, we extract phoneme representations from text input, which is then given as input to the Transformer encoder layers. Our text encoder block is similar to the one used in FastSpeech2~\cite{fastspeech2}, which consists of a positional encoding layer and Feed-Forward Transformer (FFT) layers. The phonemes are transformed to encode the semantic representation and output the text embedding vectors, $E_{text}$ of dimension: $N \times d$, where $d$ is the transformer feature dimension.

\paragraph{Visual Encoder:}
The input to our visual encoder is the visual feature sequence $g$ obtained from the lip-to-text network. We highlight that these visual features, which capture the lip shape and motion, play a crucial role in generating speech that syncs with the input video. Since these representations were also learned using text supervision, they are likely to reflect accurate content information. This starkly contrasts previous works that directly learn visual representations from speech supervision only, which can lead to sub-par visual representations that might contain other unnecessary information, such as the input face identity. The superiority of these representations is one of the critical reasons for our overall network's performance. The extracted $N \times T \times fd$ dimensional representations are given as input to the Transformer encoder layers as shown in Figure~\ref{fig:architecture}. Similar to the text encoder network, the visual encoder consists of a positional encoding followed by a series of FFT blocks. The encoder network outputs the learned visual embeddings, $E_{vis}$ of dimension: $N \times T \times d$.

\paragraph{Visual-Text Attention:}
Once we obtain the text and the visual embeddings, the next and the most important step is to find the alignment between these embeddings in time: which phoneme must be uttered when? The generated speech must take the content from the text embeddings and simultaneously, it should also temporally align (sync) it with the video frames. In order to achieve this, we employ a scaled-dot product attention~\cite{Vaswani2017AttentionIA} mechanism to learn the correspondence between text and video frames. Specifically, the visual embeddings $E_{vis}$ act as query, and the text embeddings $E_{text}$ act as keys and values. 
\begin{multline}
    Attention (Q, K, V) = Scaled-Dot \: Product\\
     Attention (E_{vis}, E_{text}, E_{text}) \in \mathbb{R}^{T \times d}   
\end{multline}
Through this attention module, the network learns the video-text temporal alignment, which synchronizes the generated speech with the input video frames. Now between the video sequence and melspectrograms, it is known that there exists a natural temporal alignment. The length of the melspectrograms is a constant $n$ times the length of the video. Thus, the attention output $A$ is up-sampled $n$ times to directly obtain the melspectrogram duration. This eliminates the need to train a separate duration predictor as done in FastSpeech2~\cite{fastspeech2}. In other words, the text-to-video alignment network already determines the duration of each phoneme in the speech output.    

\paragraph{Speaker Embedding:}
Unlike single-speaker models, we want our model to generate speech for any arbitrary speaker in the wild. Thus, our model also needs the voice input of the target speaker to generate the speech in his/her voice. We consider a random one-second audio segment of the target speaker and extract the speaker embedding vector $E_{voice}$ using the pre-trained identity network\footnote{\url{github.com/CorentinJ/Real-Time-Voice-Cloning}}. 

\paragraph{Spectrogram Decoder:}
The speaker embedding vector $E_{voice}$ is added to the upsampled attention output $A$ to obtain voice-aware content representation. The spectrogram decoder, consisting of transformer decoder layers, ingests this representation and generates the melspectrogram sequence $S_m$. To further improve speech quality, as done in most of the TTS networks, we adopt a pre-trained neural vocoder model BigVGAN~\cite{bigvgan}, to synthesize the speech from the melspectrogram output. Note that this step is only used during inference to obtain high-quality speech outputs.

\subsection{Datasets and Training Settings}

\subsubsection{Datasets}
We evaluate both constrained and unconstrained datasets to analyze the model's performance. The first corpus we experiment with is the TCD-TIMIT~\cite{harte2015tcd} lip speaker dataset, which comprises lab-recorded videos of $3$ speakers. Next, we consider the word-level LRW~\cite{chung2016lip} dataset, consisting of around $150$ hours of single-word utterances from hundreds of speakers. We then move on to the more challenging large-scale datasets: LRS2~\cite{chung2017lip} and LRS3~\cite{afouras2018lrs3}. The LRS2 data comprises thousands of speakers from BBC programs with a vocabulary of $59k$ and around $230$ hours of video clips (both ``train" and ``pre-train" sets together). The LRS3 dataset, on the other hand, is also a large-scale dataset with a total of approximately $430$ hours (``train" and ``pre-train" sets) of video data with $150k$ utterances. It consists of thousands of spoken sentences from TED and TEDx talks in English. We train and test the performance of our network using the official splits of LRW, LRS2 and LRS3 datasets and use the train-test split proposed in Lip2Wav~\cite{Prajwal_2020_CVPR} for TIMIT dataset~\cite{harte2015tcd}.

\subsubsection{Data pre-processing}
We sample the video frames at $25$ FPS and follow the pre-processing procedure of VTP~\cite{Prajwal_2022_CVPR} to obtain the face crops. For the speech segments, we compute STFT and then melspectrograms of $80$ mel-bands, with a hop length of $10ms$ and a window length of $25$ms, sampled at $16$kHz. We use an open-source grapheme-to-phoneme tool for text processing to obtain the phoneme inputs for our Visual TTS model.

\subsubsection{Model configuration and training}
Our Visual TTS model comprises $4$ FFT blocks in the text and visual encoders and $6$ FFT blocks in the spectrogram decoder network. The visual embeddings obtained from VTP are $512$-dimensional embeddings for each frame. In the video-text attention sub-network, the upsample factor, $n$ is set to $4$. For the speaker embedding, the identity network outputs a $256$ dimensional vector for each speech sample. For the lip-to-text network, we use the publicly released pre-trained model\footnote{\url{https://github.com/prajwalkr/vtp}} (trained on LRS2~\cite{chung2017lip} and LRS3~\cite{afouras2018lrs3}). Our visual TTS model is trained on a single NVIDIA $2080$ Ti GPU. We use the Adam optimizer~\cite{adam} with $\beta_1 = 0.9$, $\beta_2 = 0.98$, and $\epsilon = 10^{-9}$ and follow the same learning rate schedule as done in~\cite{Prajwal_2022_CVPR}. We set the batch size to be $16$ for all the datasets and train the model using $L_1$ reconstruction loss for approximately $900k$ steps (until convergence). We also train BigVGAN vocoder~\cite{bigvgan} and use it during inference to generate the speech from the output melspectrograms.

\section{Experiments}
We present our approach's quantitative results and comparisons with the existing methods. As automatic speech metrics are imperfect, we also show MOS scores using human evaluation. Finally, we demonstrate a real-world application of lip-to-speech for the first time by voicing the silent lip movements of an ALS patient.

\subsection{Quantitative Evaluations}

\noindent

\textbf{Metrics:}
We measure the quality of the generated speech using the standard speech metrics: Perceptual Evaluation of Speech Quality (PESQ), Short-Time Objective Intelligibility measure (STOI), and its extended version (ESTOI). PESQ measures the clarity and overall perceptual quality of speech and STOI and ESTOI measure the intelligibility of speech. Further, as discussed previously, it is very crucial to generate speech that is in sync with the input video. We use the lip-sync metrics, Lip-Sync Error - Confidence (LSE-C) and Lip-Sync Error - Distance (LSE-D)~\cite{Chung16a} to evaluate whether the output speech matches the input lip movements. We use the public implementations of all the above metrics for a fair comparison.

\subsubsection{Speech Synthesis in Constrained Settings}

\noindent

\textbf{Comparisons:}
To evaluate lip-to-speech methods on the constrained single-speaker TCD-TIMIT dataset, we compare four existing approaches: (i) GAN-based~\cite{vougioukas2019video}, (ii) Lip2Wav~\cite{Prajwal_2020_CVPR}, (iii) VAE-GAN~\cite{msl2s}, and (iv) VCA-GAN~\cite{vcagan}. We adopt the same settings as Lip2Wav~\cite{Prajwal_2020_CVPR} and report the scores~\cite{Prajwal_2020_CVPR} and VCA-GAN~\cite{vcagan}. We use the Wav2Lip~\cite{wav2lip} repository to compute the LSE-C and LSE-D scores for each method. We have excluded the metrics for a particular model that were not mentioned in the original papers or for which no publicly available pre-trained checkpoint exists.

\noindent

\textbf{Results:}
Table~\ref{table:results} contains the results on the TCD-TIMIT dataset. We observe that our approach achieves comparable results to previous methods in constrained settings with minimal data of only $3$ speakers. However, the significant benefits of our approach can be seen in unconstrained settings, which we describe below.

\begin{table*}[ht]
    \centering
    \tiny
    \caption{We compare against state-of-the-art methods on several standard multi-speaker benchmarks using standard metrics. The generated outputs from our model are most natural (PESQ), most accurate (STOI, ESTOI) and in perfect sync with the video input (LSE-C, LSE-D) in the in-the-wild videos of LRW~\cite{chung2016lip}, LRS2~\cite{chung2017lip} and LRS3~\cite{afouras2018lrs3}.}
    
    \vspace{-10pt}
    \resizebox{0.85\linewidth}{!}{
    \begin{tabular}{c|c|ccccc}
    \hline
 
   \textbf{Dataset} & \textbf{Method} & \textbf{PESQ}$\uparrow$ & \textbf{STOI}$\uparrow$ & \textbf{ESTOI}$\uparrow$ & \textbf{LSE-C}$\uparrow$ & \textbf{LSE-D}$\downarrow$ \\
    
    \hline

    \multirow{5}{*}{TCD-TIMIT~\cite{harte2015tcd}} & GAN-based~\cite{vougioukas2019video} & 1.22 & 0.51 & 0.32 & - & - \\
    & Lip2Wav~\cite{Prajwal_2020_CVPR} & 1.35 & 0.56 & 0.36 & 6.610 & 7.815 \\
    & VCA-GAN~\cite{vcagan} & \textbf{1.43} & 0.58 & 0.40 & - & - \\
    & VAE-GAN~\cite{msl2s} & 1.35 & 0.55 & 0.35 & - & - \\
    & \textbf{Ours} & 1.34 & \textbf{0.61} & \textbf{0.42} & \textbf{6.623} & \textbf{6.901} \\

    \hline
    \hline

    \multirow{8}{*}{LRW~\cite{chung2016lip}} & GAN-based~\cite{vougioukas2019video} & 0.72 & 0.10 & 0.02 & 1.983 & 9.426 \\
    & Lip2Wav~\cite{Prajwal_2020_CVPR} & 1.19 & 0.54 & 0.34 & 2.526 & 8.286 \\
    & VAE-GAN~\cite{msl2s} & 0.78 & 0.15 & 0.03 & 2.538 & 8.173 \\
    & VCA-GAN~\cite{vcagan} & 1.33 & 0.56 & 0.36 & - & - \\
    & SVTS~\cite{svts} & 1.49 & 0.64 & 0.48 & - & - \\
    & Multi-task L2S~\cite{multitask-l2s} & 1.56 & 0.64 & 0.47 & 4.876  & 8.102 \\
    & Lip-to-Text + TTS baseline & 0.69 & 0.10 & 0.01 & 1.993 & 12.872 \\
    & \textbf{Ours} & \textbf{1.61} & \textbf{0.71} & \textbf{0.56} & \textbf{6.812} & \textbf{6.974} \\
        
    \hline

    \multirow{7}{*}{LRS2~\cite{chung2017lip}} & Lip2Wav~\cite{Prajwal_2020_CVPR} & 0.58 & 0.28 & 0.11 & 1.874 & 11.48 \\
    & VAE-GAN~\cite{vougioukas2019video} & 0.60 & 0.34 & 0.17 & 2.507 & 8.155 \\
    & VCA-GAN~\cite{vcagan} & 1.24 & 0.40 & 0.13 & 4.016 & 7.914 \\
    & SVTS~\cite{svts} & 1.34 & 0.49 & 0.29 & - & - \\
    & Multi-task L2S~\cite{multitask-l2s} & 1.36 & 0.52 & 0.34 & 4.001 & 8.192 \\
    & Lip-to-Text + TTS baseline & 0.53 & 0.19 & 0.02 & 2.013 & 15.891 \\
    & \textbf{Ours} & \textbf{1.47} & \textbf{0.65} & \textbf{0.47} & \textbf{8.083} & \textbf{6.586} \\
    
    \hline

    \multirow{6}{*}{LRS3~\cite{afouras2018lrs3}} & VAE-GAN~\cite{vougioukas2019video} & 0.51 & 0.30 & 0.15 & 2.063 & 8.256 \\
    & VCA-GAN~\cite{vcagan} & 1.23 & 0.47 & 0.20 & 3.905 & 8.392 \\
    & SVTS~\cite{svts} & 1.25 & 0.50 & 0.27 & - & - \\
    & Multi-task L2S~\cite{multitask-l2s} & 1.31 & 0.48 & 0.26 & 3.876 & 8.677 \\
    & Lip-to-Text + TTS baseline & 0.42 & 0.16 & 0.01 & 1.771 & 17.882 \\
    & \textbf{Ours} & \textbf{1.39} & \textbf{0.58} & \textbf{0.37} & \textbf{7.886} & \textbf{6.850} \\

    \hline
    
    \end{tabular}
    }
    \vspace{-5pt}
    \label{table:results}
\end{table*}

\subsubsection{Speech Synthesis in Unconstrained Settings}

\noindent

\textbf{Comparisons:}
In order to assess the performance of lip-to-speech methods in unconstrained scenarios, we employ three datasets: word-level LRW~\cite{chung2016lip}, sentence-level LRS2~\cite{chung2017lip}, and LRS3~\cite{afouras2018lrs3}. While the authors of VAE-GAN~\cite{msl2s} have re-trained the GAN-based~\cite{vougioukas2019video} and Lip2Wav~\cite{Prajwal_2020_CVPR} models in a multi-speaker context, we present the scores from their original study for comparison. For VCA-GAN~\cite{vcagan}, SVTS~\cite{svts}, and Multi-task Lip-to-Speech synthesis~\cite{multitask-l2s}, we adopt the speech metric (PESQ, STOI and ESTOI) scores from~\cite{multitask-l2s}. Further, we use publicly accessible pre-trained checkpoints for VCA-GAN\footnote{\url{https://github.com/ms-dot-k/Visual-Context-Attentional-GAN}} and Multitask-L2S\footnote{\url{https://github.com/ms-dot-k/Lip-to-Speech-Synthesis-in-the-Wild}} to generate speech on LRS2 and LRS3 test sets for the former, and LRS2, LRS3, and LRW test sets for the latter. We utilize these generations to compute the LSE metrics for both techniques, wherever applicable. Lastly, we include results for a baseline approach that leverages lip-to-text conversion followed by multi-speaker TTS without a visual stream in the TTS model. Our evaluation does not include all metrics that were not originally reported in the papers or for which no pre-trained model is publicly available.

\noindent

\begin{figure}[h]
  \centering
  \vspace{-10pt}
  \includegraphics[width=\linewidth]{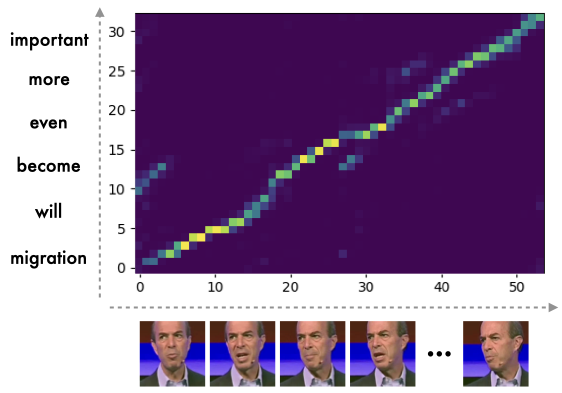}
  \vspace{-15pt}
  \caption{We visualize the video-text alignment from the scaled dot product attention step of our model. We observe that the model learns a strong monotonic near-diagonal attention, as expected.}
  \label{fig:attn_plot}
  \vspace{-10pt}
\end{figure}

\textbf{Results:}
We present the results on the challenging LRW, LRS2, and LRS3 datasets in Table~\ref{table:results}. Our model consistently outperforms the existing methods by a significant margin on all these datasets. Since GAN-based~\cite{vougioukas2019video} model was proposed to work for constrained laboratory recorded datasets, we can observe that extending this model in unconstrained settings does not yield satisfactory results. Lip2Wav~\cite{Prajwal_2020_CVPR} performs decently on the word-level LRW dataset; however, it fails to learn the audio-visual alignment on the LRS2 dataset, thus leading to very poor performance. We discard this model for further comparison on the LRS3 dataset. VAE-GAN~\cite{msl2s}, VCA-GAN~\cite{vcagan}, SVTS~\cite{svts} and Multitask-L2S~\cite{multitask-l2s} generate speech that is in-sync with the input video; however, they fail to synthesize accurate content. The quality of the generated speech is often non-intelligible and leads to lower scores in speech quality metrics. The Lip-to-Text $+$ TTS baseline model is on the opposite spectrum, where the model generates the content well but fails to capture the lip-sync, mainly because the model cannot infer the speed, prosody, and accents of speakers just from the text input. Our model, on the other hand, is capable of generating both the actual spoken content as well as maintaining precise lip synchronization. As we can see from the table, we outperform the previous methods in all the speech quality metrics, indicating the robustness and superiority of our approach. In Figure~\ref{fig:attn_plot}, we depict how our model temporally aligns video and text sequences in the process of generating speech. We encourage the reader to view our demo video comprising multiple qualitative samples and comparisons.  

\subsection{Human Evaluations}
To evaluate the applicability of our method in real-world scenarios, we conduct subjective human evaluations. We ask $25$ volunteers to assess the quality of speech generations. The participant group has an almost equal male-female ratio, spanning an age group of $20 - 45$ years. We randomly select $10$ long sentences ($10$ seconds or longer) from the test set of LRS3~\cite{afouras2018lrs3} and present the results from different methods to the participants. We ask them to rate the samples on a scale of $1 - 5$ based on the following criteria: (A) Intelligibility (is the speech meaningful?), (B) Content clarity (are the words clear?), (C) Sync Accuracy, (D) Overall perceptual quality of the talking head video $+$ audio. We report the mean opinion scores in Table~\ref{table:human_evals}. On par with the quantitative evaluations, our method is highly rated over other approaches in all the criteria listed above. As expected, the speech intelligibility is slightly rated higher for the Lip-to-Text $+$ TTS baseline. However, the overall perceptual quality for this baseline sharply falls due to the lack of sync between the spoken content and the lip movements. Overall, Table~\ref{table:human_evals} clearly signifies that our network is able to generate speech with more clarity, which sounds more natural and is of considerably higher quality. 

\begin{table}[t]
    \centering
    \caption{(A) Intelligibility, (B) Content clarity, (C) Sync Accuracy, (D) Overall perceptual quality. Our model produces natural and realistic speech outputs that is largely preferred by the users in comparison to other approaches.}
    
    \vspace{-10pt}
    \begin{tabular}{lcccc}
    \hline

    Method & (A) & (B) & (C) & (D) \\ \hline 
    
    GAN-based~\cite{vougioukas2019video} & 2.05 & 1.87 & 1.99 & 2.12 \\
    Lip2Wav~\cite{Prajwal_2020_CVPR} & 1.01 & 1.03 & 1.34 & 1.01 \\
    VAE-GAN~\cite{msl2s} & 1.07 & 1.33 & 2.18 & 2.57\\
    VCA-GAN~\cite{vcagan} & 2.18 & 1.88 & 2.97 & 2.54 \\
    Multi-task L2S~\cite{multitask-l2s} & 2.19 & 1.85 & 3.01 & 2.64\\
    Lip-to-Text + TTS baseline & \textbf{3.61} & 2.87 & 1.01 & 2.96 \\
            
    \textbf{Ours} & 3.49 & \textbf{3.52} & \textbf{3.82} & \textbf{3.31}\\
    
    \hline
    
    \end{tabular}
    \vspace{-10pt}
    \label{table:human_evals}
\end{table}

\vspace{-8pt}
\subsection{Applications in Assistive Technology}
\label{sec:applications_l2s}
Lip-to-Speech synthesis has a host of applications in a world that is becoming increasingly digital. Simple applications such as performing video calls in quiet environments, filling in the audio interruptions due to technical issues, eliminating unwanted background chatter, etc., can be made possible with accurate lip-to-speech. We believe the most significant application of lip-to-speech can be in assistive technologies. It can revolutionize the current assistive systems used to improve the communication ability of people suffering from various disorders affecting their speech. Patients suffering from vocal cord disabilities can mouth words to communicate naturally with the world around them. The synthesized speech can be personalized and also be in sync with the speaker's lip movements. 

\paragraph{Generating Speech for a Patient suffering from ALS:}
A recent work~\cite{als-lr} proposed a lip reading technique for a patient suffering from ALS. The patient has feeble vocal cord movements but can mouth words silently. The authors of the paper collected limited amounts of data from the patient and trained a model to recognize words and sentences from his lip movements. As a result of our significant improvement in this task, we can now demonstrate the benefit of using lip-to-speech as a future assistive technology. Through the help of authors of~\cite{als-lr}, we evaluate our lip-to-speech system on the patients' data. Please note that the data was anonymized and was used only for research purposes. We find that the lip-to-text module generates fairly accurate text (WER of $\approx37\%$), and the visual TTS model generates clear speech in sync with the patient's lip movements. This is the first demonstration of automatic lip-to-speech synthesis for an unseen speaker in an entirely out-of-domain real-world application. Further, we also test our model on the other deaf speakers studied in~\cite{als-lr} and observe accurate performance.

\begin{figure}[h]
  \centering
  \includegraphics[width=\linewidth]{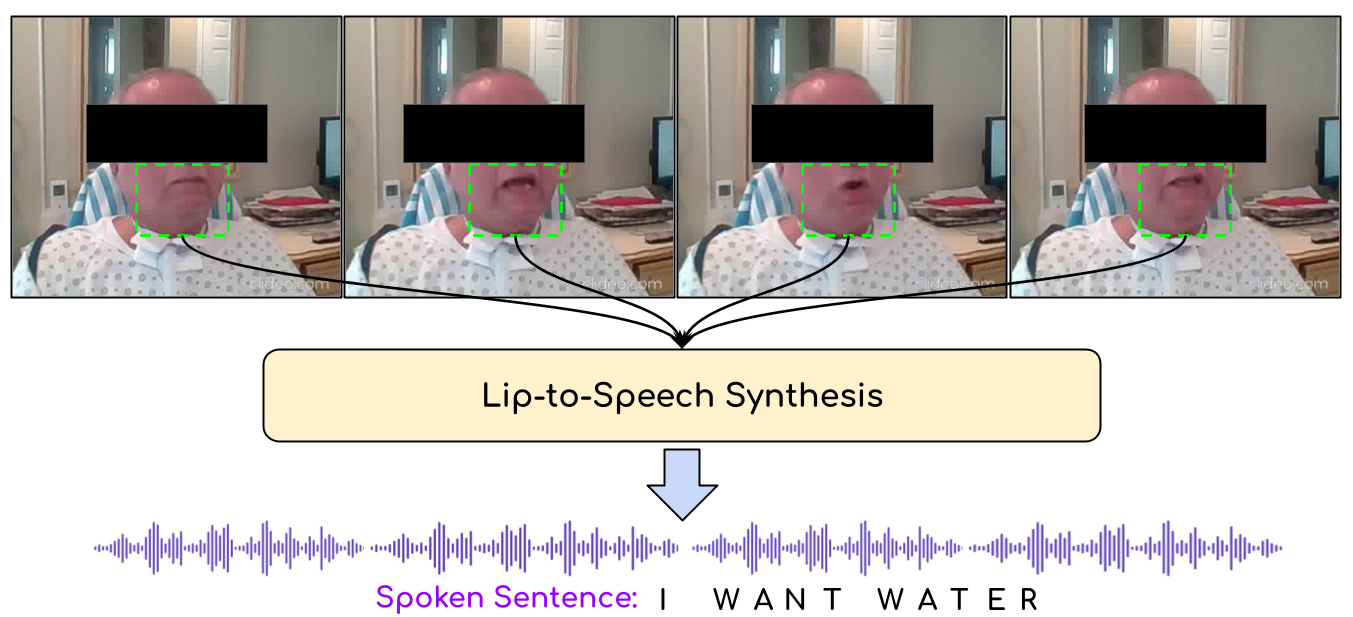}
  \vspace{-10pt}
  \caption{We demonstrate our model on an ALS patient who cannot voice words but can mouth them. We can generate the speech corresponding to the silent lip movements. Lip-to-Speech can thus be a cheap and non-invasive method to assist someone who has lost their voice.}
  \label{fig:als}
  \vspace{-10pt}
\end{figure}

\paragraph{Ethical Considerations}
We acknowledge that our work has the potential to generate synthetic speech for videos, given that we only require a $1-$second voice sample from any target speaker. However, since the video provides context and constrains the output speech, the generated speech will likely follow the original content closely. We recognize the importance of ethical considerations regarding using such models and ensure that our models will only be shared with users who consent to limit their usage to research-oriented and ethically valid tasks.

\section{Ablation Studies}
In this section, we perform several ablation studies to understand the effect of different components of our model. All the ablation experiments are performed on the LRS2~\cite{chung2017lip} test set.

\paragraph{Effect of different pre-trained lip-to-text models}
Additional experiments were conducted using other pre-trained lip-to-text models, specifically DeepLR~\cite{Afouras2018DeepLR} and AV-HuBERT~\cite{av-hubert}. The former had a WER of $51.3$ on the LRS2 test set, while the latter had a WER of $46.6$. 
The input text to our pre-trained Visual TTS module was taken from different Lip-to-Text models. Additionally, we also directly provide the ground-truth text from the LRS2 test set. As shown in Table~\ref{table:lipreading}, our approach recovered speech that was somewhat accurate, despite the presence of noisy text transcripts from both models. This can be attributed to two factors: (i) the correction of errors made by the lip-to-text network by our pipeline to some extent (demonstrated in the demo video); and (ii) the reduced difference in scores between homonyms such as "ship and sheep" or "berth and birth" in the audio domain.

\begin{table}[ht]
    \centering
    \small
    \setlength{\tabcolsep}{1pt}
    \caption{Comparison of using generated text from different lip-to-text network in our pipeline. We also report the WER of the lip reading model (L2T-WER) on the LRS2 test set as a reference.}
    \vspace{-10pt}
    \begin{tabular}{l|ccccccccc}
    \hline

    \textbf{Method} & \textbf{L2T-WER} &
    \textbf{PESQ}$\uparrow$ & \textbf{STOI}$\uparrow$ & \textbf{ESTOI}$\uparrow$ & \textbf{LSE-C}$\uparrow$ & \textbf{LSE-D}$\downarrow$\\
    \hline
    
    Deep Lip Reading~\cite{Afouras2018DeepLR} & 51.3 & 1.17 & 0.40 & 0.22 & 7.847 & 6.904\\
    AV-HuBERT~\cite{av-hubert} & 46.1 & 1.27 & 0.53 & 0.40 & 7.960 & 7.003\\
    VTP (Ours) & 22.6 & 1.47 & 0.65 & 0.47 & 8.083 & 6.586 \\
    GT text & - & 1.51 & 0.69 & 0.50 & 8.781 & 6.106\\
        
    \hline
    
    \end{tabular}
    \vspace{-8pt}
    \label{table:lipreading}
\end{table} 

\paragraph{Effect of different visual representations}
We train our proposed Visual Text-to-Speech module with RGB face crops instead of the VTP embeddings to generate speech conditioned on text and lip movements. Based on our observations from Table~\ref{tab:visual_representations}, VTP embeddings are the most suitable for this task because they excel in localizing and representing the shape of the speaker's lips.

\begin{table}[ht]
    \centering
    \setlength{\tabcolsep}{3pt}
    \vspace{-5pt}
    \caption{We present the effect of using different visual representations for training the Visual TTS module.}
    \vspace{-10pt}
    \begin{tabular}{l|cccccccc}
    \hline

    \textbf{Method}  &
    \textbf{PESQ}$\uparrow$ & \textbf{STOI}$\uparrow$ & \textbf{ESTOI}$\uparrow$ & \textbf{LSE-C}$\uparrow$ & \textbf{LSE-D}$\downarrow$\\
    \hline
    
    Face crops & 1.17 & 0.40 & 0.22 & 7.847 & 6.904\\
    VTP (Ours) & 1.47 & 0.65 & 0.47 & 8.083 & 6.586\\
        
    \hline
    
    \end{tabular}
    \vspace{-10pt}
    \label{tab:visual_representations}
\end{table}

\section{Limitations}
In our work, we address the problem of lip-to-speech networks not being able to learn a language model directly from speech supervision. We do so by using a pre-trained lip-to-text network. While our model does not require ground-truth text annotations, the lip-to-text model which we build upon has been trained with text supervision. However, recent efforts in self-supervised pre-training have led to a sharp decrease in the number of text annotations required for training accurate lip-to-text~\cite{av-hubert}, making it easier to extend such models to lip-to-speech using our approach. Currently, we have only tested our model in English, and it remains to be validated in other languages. 

\section{Conclusion}
Our research presents an innovative approach to unconstrained multi-speaker lip-to-speech synthesis that outperforms previous methods by incorporating language and visual information from a highly accurate lip-to-text model. We demonstrate significant improvements in lip-to-speech synthesis, generating high-quality outputs that seamlessly synchronize with silent lip video. Our study can potentially open up exciting avenues for future research. We are particularly encouraged by the success of our approach in assistive technology, where we have shown that our method can generate accurate speech from silent lip movements of individuals with speech impairments. Overall, we are optimistic about the possibilities of our approach to improve communication and enhance the quality of life for people with speech impairments, and we look forward to seeing our work drive further progress in this field.

\noindent

\textbf{Acknowledgement:} This work is supported by MeitY, Government of India

\bibliographystyle{ACM-Reference-Format}
\balance
\bibliography{acmart}


\begin{thebibliography}{36}


\ifx \showCODEN    \undefined \def \showCODEN     #1{\unskip}     \fi
\ifx \showDOI      \undefined \def \showDOI       #1{#1}\fi
\ifx \showISBNx    \undefined \def \showISBNx     #1{\unskip}     \fi
\ifx \showISBNxiii \undefined \def \showISBNxiii  #1{\unskip}     \fi
\ifx \showISSN     \undefined \def \showISSN      #1{\unskip}     \fi
\ifx \showLCCN     \undefined \def \showLCCN      #1{\unskip}     \fi
\ifx \shownote     \undefined \def \shownote      #1{#1}          \fi
\ifx \showarticletitle \undefined \def \showarticletitle #1{#1}   \fi
\ifx \showURL      \undefined \def \showURL       {\relax}        \fi
\providecommand\bibfield[2]{#2}
\providecommand\bibinfo[2]{#2}
\providecommand\natexlab[1]{#1}
\providecommand\showeprint[2][]{arXiv:#2}

\bibitem[Afouras et~al\mbox{.}(2018c)]%
        {DeepAVLR_tpami_2018}
\bibfield{author}{\bibinfo{person}{Triantafyllos Afouras},
  \bibinfo{person}{Joon~Son Chung}, \bibinfo{person}{Andrew Senior},
  \bibinfo{person}{Oriol Vinyals}, {and} \bibinfo{person}{Andrew Zisserman}.}
  \bibinfo{year}{2018}\natexlab{c}.
\newblock \showarticletitle{Deep audio-visual speech recognition}.
\newblock \bibinfo{journal}{\emph{IEEE transactions on pattern analysis and
  machine intelligence}} (\bibinfo{year}{2018}).
\newblock


\bibitem[Afouras et~al\mbox{.}(2018a)]%
        {Afouras2018DeepLR}
\bibfield{author}{\bibinfo{person}{Triantafyllos Afouras},
  \bibinfo{person}{Joon~Son Chung}, {and} \bibinfo{person}{Andrew Zisserman}.}
  \bibinfo{year}{2018}\natexlab{a}.
\newblock \showarticletitle{Deep Lip Reading: a comparison of models and an
  online application}. In \bibinfo{booktitle}{\emph{INTERSPEECH}}.
\newblock


\bibitem[Afouras et~al\mbox{.}(2018b)]%
        {afouras2018lrs3}
\bibfield{author}{\bibinfo{person}{Triantafyllos Afouras},
  \bibinfo{person}{Joon~Son Chung}, {and} \bibinfo{person}{Andrew Zisserman}.}
  \bibinfo{year}{2018}\natexlab{b}.
\newblock \showarticletitle{LRS3-TED: a large-scale dataset for visual speech
  recognition}.
\newblock \bibinfo{journal}{\emph{arXiv preprint arXiv:1809.00496}}
  (\bibinfo{year}{2018}).
\newblock


\bibitem[Akbari et~al\mbox{.}(2017)]%
        {Akbari2017Lip2AudspecSR}
\bibfield{author}{\bibinfo{person}{Hassan Akbari}, \bibinfo{person}{Himani
  Arora}, \bibinfo{person}{Liangliang Cao}, {and} \bibinfo{person}{Nima
  Mesgarani}.} \bibinfo{year}{2017}\natexlab{}.
\newblock \showarticletitle{Lip2Audspec: Speech Reconstruction from Silent Lip
  Movements Video}.
\newblock \bibinfo{journal}{\emph{2018 IEEE International Conference on
  Acoustics, Speech and Signal Processing (ICASSP)}} (\bibinfo{year}{2017}),
  \bibinfo{pages}{2516--2520}.
\newblock


\bibitem[Assael et~al\mbox{.}(2016)]%
        {LipNet_arxiv_2016}
\bibfield{author}{\bibinfo{person}{Yannis~M Assael}, \bibinfo{person}{Brendan
  Shillingford}, \bibinfo{person}{Shimon Whiteson}, {and}
  \bibinfo{person}{Nando De~Freitas}.} \bibinfo{year}{2016}\natexlab{}.
\newblock \showarticletitle{Lipnet: End-to-end sentence-level lipreading}.
\newblock \bibinfo{journal}{\emph{arXiv preprint arXiv:1611.01599}}
  (\bibinfo{year}{2016}).
\newblock


\bibitem[Chan et~al\mbox{.}(2016)]%
        {las}
\bibfield{author}{\bibinfo{person}{William Chan}, \bibinfo{person}{Navdeep
  Jaitly}, \bibinfo{person}{Quoc Le}, {and} \bibinfo{person}{Oriol Vinyals}.}
  \bibinfo{year}{2016}\natexlab{}.
\newblock \showarticletitle{Listen, attend and spell: A neural network for
  large vocabulary conversational speech recognition}. In
  \bibinfo{booktitle}{\emph{2016 IEEE International Conference on Acoustics,
  Speech and Signal Processing (ICASSP)}}. \bibinfo{pages}{4960--4964}.
\newblock
\urldef\tempurl%
\url{https://doi.org/10.1109/ICASSP.2016.7472621}
\showDOI{\tempurl}


\bibitem[Chung et~al\mbox{.}(2017)]%
        {chung2017lip}
\bibfield{author}{\bibinfo{person}{Joon~Son Chung}, \bibinfo{person}{Andrew
  Senior}, \bibinfo{person}{Oriol Vinyals}, {and} \bibinfo{person}{Andrew
  Zisserman}.} \bibinfo{year}{2017}\natexlab{}.
\newblock \showarticletitle{Lip reading sentences in the wild}. In
  \bibinfo{booktitle}{\emph{2017 IEEE Conference on Computer Vision and Pattern
  Recognition (CVPR)}}. IEEE, \bibinfo{pages}{3444--3453}.
\newblock


\bibitem[Chung and Zisserman(2016a)]%
        {chung2016lip}
\bibfield{author}{\bibinfo{person}{Joon~Son Chung} {and}
  \bibinfo{person}{Andrew Zisserman}.} \bibinfo{year}{2016}\natexlab{a}.
\newblock \showarticletitle{Lip reading in the wild}. In
  \bibinfo{booktitle}{\emph{Asian Conference on Computer Vision}}. Springer,
  \bibinfo{pages}{87--103}.
\newblock


\bibitem[Chung and Zisserman(2016b)]%
        {Chung16a}
\bibfield{author}{\bibinfo{person}{J.~S. Chung} {and} \bibinfo{person}{A.
  Zisserman}.} \bibinfo{year}{2016}\natexlab{b}.
\newblock \showarticletitle{Out of time: automated lip sync in the wild}. In
  \bibinfo{booktitle}{\emph{Workshop on Multi-view Lip-reading, ACCV}}.
\newblock


\bibitem[Cooke et~al\mbox{.}(2006)]%
        {cooke2006audio}
\bibfield{author}{\bibinfo{person}{Martin Cooke}, \bibinfo{person}{Jon Barker},
  \bibinfo{person}{Stuart Cunningham}, {and} \bibinfo{person}{Xu Shao}.}
  \bibinfo{year}{2006}\natexlab{}.
\newblock \showarticletitle{An audio-visual corpus for speech perception and
  automatic speech recognition}.
\newblock \bibinfo{journal}{\emph{The Journal of the Acoustical Society of
  America}} \bibinfo{volume}{120}, \bibinfo{number}{5} (\bibinfo{year}{2006}),
  \bibinfo{pages}{2421--2424}.
\newblock


\bibitem[Ephrat et~al\mbox{.}(2017)]%
        {Ephrat2017ImprovedSR}
\bibfield{author}{\bibinfo{person}{Ariel Ephrat}, \bibinfo{person}{Tavi
  Halperin}, {and} \bibinfo{person}{Shmuel Peleg}.}
  \bibinfo{year}{2017}\natexlab{}.
\newblock \showarticletitle{Improved Speech Reconstruction from Silent Video}.
\newblock \bibinfo{journal}{\emph{2017 IEEE International Conference on
  Computer Vision Workshops (ICCVW)}} (\bibinfo{year}{2017}),
  \bibinfo{pages}{455--462}.
\newblock


\bibitem[Ephrat and Peleg(2017)]%
        {ephrat2017vid2speech}
\bibfield{author}{\bibinfo{person}{Ariel Ephrat} {and} \bibinfo{person}{Shmuel
  Peleg}.} \bibinfo{year}{2017}\natexlab{}.
\newblock \showarticletitle{Vid2Speech: speech reconstruction from silent
  video}. In \bibinfo{booktitle}{\emph{2017 IEEE International Conference on
  Acoustics, Speech and Signal Processing (ICASSP)}}.
\newblock


\bibitem[Freitag and Al-Onaizan(2017)]%
        {beam}
\bibfield{author}{\bibinfo{person}{Markus Freitag} {and} \bibinfo{person}{Yaser
  Al-Onaizan}.} \bibinfo{year}{2017}\natexlab{}.
\newblock \showarticletitle{Beam Search Strategies for Neural Machine
  Translation}. In \bibinfo{booktitle}{\emph{NMT@ACL}}.
\newblock


\bibitem[gil Lee et~al\mbox{.}(2022)]%
        {bigvgan}
\bibfield{author}{\bibinfo{person}{Sang gil Lee}, \bibinfo{person}{Wei Ping},
  \bibinfo{person}{Boris Ginsburg}, \bibinfo{person}{Bryan Catanzaro}, {and}
  \bibinfo{person}{Sung-Hoon Yoon}.} \bibinfo{year}{2022}\natexlab{}.
\newblock \showarticletitle{BigVGAN: A Universal Neural Vocoder with
  Large-Scale Training}.
\newblock \bibinfo{journal}{\emph{ArXiv}}  \bibinfo{volume}{abs/2206.04658}
  (\bibinfo{year}{2022}).
\newblock


\bibitem[Harte and Gillen(2015)]%
        {harte2015tcd}
\bibfield{author}{\bibinfo{person}{Naomi Harte} {and} \bibinfo{person}{Eoin
  Gillen}.} \bibinfo{year}{2015}\natexlab{}.
\newblock \showarticletitle{TCD-TIMIT: An audio-visual corpus of continuous
  speech}.
\newblock \bibinfo{journal}{\emph{IEEE Transactions on Multimedia}}
  \bibinfo{volume}{17}, \bibinfo{number}{5} (\bibinfo{year}{2015}),
  \bibinfo{pages}{603--615}.
\newblock


\bibitem[Hassid et~al\mbox{.}(2022)]%
        {vdtts}
\bibfield{author}{\bibinfo{person}{Michael Hassid},
  \bibinfo{person}{Michelle~Tadmor Ramanovich}, \bibinfo{person}{Brendan
  Shillingford}, \bibinfo{person}{Miaosen Wang}, \bibinfo{person}{Ye Jia},
  {and} \bibinfo{person}{Tal Remez}.} \bibinfo{year}{2022}\natexlab{}.
\newblock \showarticletitle{More than Words: In-the-Wild Visually-Driven
  Prosody for Text-to-Speech}.
\newblock \bibinfo{journal}{\emph{2022 IEEE/CVF Conference on Computer Vision
  and Pattern Recognition (CVPR)}} (\bibinfo{year}{2022}),
  \bibinfo{pages}{10577--10587}.
\newblock


\bibitem[Hegde et~al\mbox{.}(2022)]%
        {msl2s}
\bibfield{author}{\bibinfo{person}{Sindhu~B. Hegde}, \bibinfo{person}{K~R
  Prajwal}, \bibinfo{person}{Rudrabha Mukhopadhyay}, \bibinfo{person}{Vinay~P.
  Namboodiri}, {and} \bibinfo{person}{C.V. Jawahar}.}
  \bibinfo{year}{2022}\natexlab{}.
\newblock \showarticletitle{Lip-to-Speech Synthesis for Arbitrary Speakers in
  the Wild}. In \bibinfo{booktitle}{\emph{Proceedings of the 30th ACM
  International Conference on Multimedia}} (Lisboa, Portugal)
  \emph{(\bibinfo{series}{MM '22})}. \bibinfo{publisher}{Association for
  Computing Machinery}, \bibinfo{address}{New York, NY, USA},
  \bibinfo{pages}{6250–6258}.
\newblock
\showISBNx{9781450392037}
\urldef\tempurl%
\url{https://doi.org/10.1145/3503161.3548081}
\showDOI{\tempurl}


\bibitem[Hu et~al\mbox{.}(2021)]%
        {neural_dubber}
\bibfield{author}{\bibinfo{person}{Chenxu Hu}, \bibinfo{person}{Qiao Tian},
  \bibinfo{person}{Tingle Li}, \bibinfo{person}{Yuping Wang},
  \bibinfo{person}{Yuxuan Wang}, {and} \bibinfo{person}{Hang Zhao}.}
  \bibinfo{year}{2021}\natexlab{}.
\newblock \showarticletitle{Neural Dubber: Dubbing for Videos According to
  Scripts}. In \bibinfo{booktitle}{\emph{NeurIPS}}.
\newblock


\bibitem[Jia et~al\mbox{.}(2018)]%
        {sv2tts}
\bibfield{author}{\bibinfo{person}{Ye Jia}, \bibinfo{person}{Yu Zhang},
  \bibinfo{person}{Ron~J. Weiss}, \bibinfo{person}{Quan Wang},
  \bibinfo{person}{Jonathan Shen}, \bibinfo{person}{Fei Ren},
  \bibinfo{person}{Zhifeng Chen}, \bibinfo{person}{Patrick Nguyen},
  \bibinfo{person}{Ruoming Pang}, \bibinfo{person}{Ignacio~Lopez Moreno}, {and}
  \bibinfo{person}{Yonghui Wu}.} \bibinfo{year}{2018}\natexlab{}.
\newblock \showarticletitle{Transfer Learning from Speaker Verification to
  Multispeaker Text-to-Speech Synthesis}. In
  \bibinfo{booktitle}{\emph{Proceedings of the 32nd International Conference on
  Neural Information Processing Systems}} \emph{(\bibinfo{series}{NIPS'18})}.
  \bibinfo{publisher}{Curran Associates Inc.}, \bibinfo{pages}{4485–4495}.
\newblock


\bibitem[Kim et~al\mbox{.}(2022)]%
        {vcagan}
\bibfield{author}{\bibinfo{person}{Minsu Kim}, \bibinfo{person}{Joanna Hong},
  {and} \bibinfo{person}{Yong~Man Ro}.} \bibinfo{year}{2022}\natexlab{}.
\newblock \showarticletitle{Lip to Speech Synthesis with Visual Context
  Attentional GAN}. In \bibinfo{booktitle}{\emph{Neural Information Processing
  Systems}}.
\newblock


\bibitem[Kim et~al\mbox{.}(2023)]%
        {multitask-l2s}
\bibfield{author}{\bibinfo{person}{Minsu Kim}, \bibinfo{person}{Joanna Hong},
  {and} \bibinfo{person}{Yong~Man Ro}.} \bibinfo{year}{2023}\natexlab{}.
\newblock \showarticletitle{Lip-to-Speech Synthesis in the Wild with Multi-task
  Learning}.
\newblock \bibinfo{journal}{\emph{ArXiv}}  \bibinfo{volume}{abs/2302.08841}
  (\bibinfo{year}{2023}).
\newblock


\bibitem[Kingma and Ba(2015)]%
        {adam}
\bibfield{author}{\bibinfo{person}{Diederik~P. Kingma} {and}
  \bibinfo{person}{Jimmy Ba}.} \bibinfo{year}{2015}\natexlab{}.
\newblock \showarticletitle{Adam: A Method for Stochastic Optimization}.
\newblock \bibinfo{journal}{\emph{CoRR}}  \bibinfo{volume}{abs/1412.6980}
  (\bibinfo{year}{2015}).
\newblock


\bibitem[Mira et~al\mbox{.}(2022)]%
        {svts}
\bibfield{author}{\bibinfo{person}{Rodrigo Mira}, \bibinfo{person}{Alexandros
  Haliassos}, \bibinfo{person}{Stavros Petridis}, \bibinfo{person}{Bj{\"o}rn
  Schuller}, {and} \bibinfo{person}{Maja Pantic}.}
  \bibinfo{year}{2022}\natexlab{}.
\newblock \showarticletitle{SVTS: Scalable Video-to-Speech Synthesis}. In
  \bibinfo{booktitle}{\emph{Interspeech}}.
\newblock


\bibitem[Momeni et~al\mbox{.}(2022)]%
        {sign_language}
\bibfield{author}{\bibinfo{person}{Liliane Momeni}, \bibinfo{person}{Hannah
  Bull}, \bibinfo{person}{Prajwal~K R}, \bibinfo{person}{Samuel Albanie},
  \bibinfo{person}{G{\"u}l Varol}, {and} \bibinfo{person}{Andrew Zisserman}.}
  \bibinfo{year}{2022}\natexlab{}.
\newblock \showarticletitle{Automatic dense annotation of large-vocabulary sign
  language videos}. In \bibinfo{booktitle}{\emph{ECCV}}.
\newblock


\bibitem[Ping et~al\mbox{.}(2018)]%
        {deepvoice3}
\bibfield{author}{\bibinfo{person}{Wei Ping}, \bibinfo{person}{Kainan Peng},
  \bibinfo{person}{Andrew Gibiansky}, \bibinfo{person}{Sercan~O. Arik},
  \bibinfo{person}{Ajay Kannan}, \bibinfo{person}{Sharan Narang},
  \bibinfo{person}{Jonathan Raiman}, {and} \bibinfo{person}{John Miller}.}
  \bibinfo{year}{2018}\natexlab{}.
\newblock \showarticletitle{Deep Voice 3: 2000-Speaker Neural Text-to-Speech}.
  In \bibinfo{booktitle}{\emph{International Conference on Learning
  Representations}}.
\newblock
\urldef\tempurl%
\url{https://openreview.net/forum?id=HJtEm4p6Z}
\showURL{%
\tempurl}


\bibitem[Prajwal et~al\mbox{.}(2022)]%
        {Prajwal_2022_CVPR}
\bibfield{author}{\bibinfo{person}{K~R Prajwal}, \bibinfo{person}{Triantafyllos
  Afouras}, {and} \bibinfo{person}{Andrew Zisserman}.}
  \bibinfo{year}{2022}\natexlab{}.
\newblock \showarticletitle{Sub-Word Level Lip Reading With Visual Attention}.
  In \bibinfo{booktitle}{\emph{Proceedings of the IEEE/CVF Conference on
  Computer Vision and Pattern Recognition (CVPR)}}.
  \bibinfo{pages}{5162--5172}.
\newblock


\bibitem[Prajwal et~al\mbox{.}(2020a)]%
        {Prajwal_2020_CVPR}
\bibfield{author}{\bibinfo{person}{K~R Prajwal}, \bibinfo{person}{Rudrabha
  Mukhopadhyay}, \bibinfo{person}{Vinay~P. Namboodiri}, {and}
  \bibinfo{person}{C.V. Jawahar}.} \bibinfo{year}{2020}\natexlab{a}.
\newblock \showarticletitle{Learning Individual Speaking Styles for Accurate
  Lip to Speech Synthesis}. In \bibinfo{booktitle}{\emph{The IEEE/CVF
  Conference on Computer Vision and Pattern Recognition (CVPR)}}.
\newblock


\bibitem[Prajwal et~al\mbox{.}(2020b)]%
        {wav2lip}
\bibfield{author}{\bibinfo{person}{K~R Prajwal}, \bibinfo{person}{Rudrabha
  Mukhopadhyay}, \bibinfo{person}{Vinay~P. Namboodiri}, {and}
  \bibinfo{person}{C.V. Jawahar}.} \bibinfo{year}{2020}\natexlab{b}.
\newblock \showarticletitle{A Lip Sync Expert Is All You Need for Speech to Lip
  Generation In the Wild}. In \bibinfo{booktitle}{\emph{Proceedings of the 28th
  ACM International Conference on Multimedia}} (Seattle, WA, USA)
  \emph{(\bibinfo{series}{MM '20})}. \bibinfo{publisher}{Association for
  Computing Machinery}, \bibinfo{address}{New York, NY, USA},
  \bibinfo{pages}{484–492}.
\newblock
\showISBNx{9781450379885}
\urldef\tempurl%
\url{https://doi.org/10.1145/3394171.3413532}
\showDOI{\tempurl}


\bibitem[R et~al\mbox{.}(2021)]%
        {KR2021VisualKS}
\bibfield{author}{\bibinfo{person}{Prajwal~K R}, \bibinfo{person}{Liliane
  Momeni}, \bibinfo{person}{Triantafyllos Afouras}, {and}
  \bibinfo{person}{Andrew Zisserman}.} \bibinfo{year}{2021}\natexlab{}.
\newblock \showarticletitle{Visual Keyword Spotting with Attention}. In
  \bibinfo{booktitle}{\emph{BMVC}}.
\newblock


\bibitem[Ren et~al\mbox{.}(2020)]%
        {fastspeech2}
\bibfield{author}{\bibinfo{person}{Yi Ren}, \bibinfo{person}{Chenxu Hu},
  \bibinfo{person}{Xu Tan}, \bibinfo{person}{Tao Qin}, \bibinfo{person}{Sheng
  Zhao}, \bibinfo{person}{Zhou Zhao}, {and} \bibinfo{person}{Tie-Yan Liu}.}
  \bibinfo{year}{2020}\natexlab{}.
\newblock \showarticletitle{Fastspeech 2: Fast and high-quality end-to-end text
  to speech}.
\newblock \bibinfo{journal}{\emph{arXiv preprint arXiv:2006.04558}}
  (\bibinfo{year}{2020}).
\newblock


\bibitem[Sen et~al\mbox{.}(2021)]%
        {als-lr}
\bibfield{author}{\bibinfo{person}{Bipasha Sen}, \bibinfo{person}{Aditya
  Agarwal}, \bibinfo{person}{Rudrabha Mukhopadhyay}, \bibinfo{person}{Vinay
  Namboodiri}, {and} \bibinfo{person}{CV Jawahar}.}
  \bibinfo{year}{2021}\natexlab{}.
\newblock \showarticletitle{Personalized One-Shot Lipreading for an ALS
  Patient}.
\newblock \bibinfo{journal}{\emph{arXiv preprint arXiv:2111.01740}}
  (\bibinfo{year}{2021}).
\newblock


\bibitem[Serdyuk et~al\mbox{.}(2022)]%
        {serdyuk22_interspeech}
\bibfield{author}{\bibinfo{person}{Dmitriy Serdyuk}, \bibinfo{person}{Otavio
  Braga}, {and} \bibinfo{person}{Olivier Siohan}.}
  \bibinfo{year}{2022}\natexlab{}.
\newblock \showarticletitle{{Transformer-Based Video Front-Ends for
  Audio-Visual Speech Recognition for Single and Muti-Person Video}}. In
  \bibinfo{booktitle}{\emph{Proc. Interspeech 2022}}.
  \bibinfo{pages}{2833--2837}.
\newblock
\urldef\tempurl%
\url{https://doi.org/10.21437/Interspeech.2022-10920}
\showDOI{\tempurl}


\bibitem[Shen et~al\mbox{.}(2018)]%
        {tacotron2}
\bibfield{author}{\bibinfo{person}{Jonathan Shen}, \bibinfo{person}{R. Pang},
  \bibinfo{person}{Ron~J. Weiss}, \bibinfo{person}{M. Schuster},
  \bibinfo{person}{Navdeep Jaitly}, \bibinfo{person}{Z. Yang},
  \bibinfo{person}{Z. Chen}, \bibinfo{person}{Yu Zhang},
  \bibinfo{person}{Yuxuan Wang}, \bibinfo{person}{R. Skerry-Ryan},
  \bibinfo{person}{R.~A. Saurous}, \bibinfo{person}{Yannis Agiomyrgiannakis},
  {and} \bibinfo{person}{Y. Wu}.} \bibinfo{year}{2018}\natexlab{}.
\newblock \showarticletitle{Natural TTS Synthesis by Conditioning Wavenet on
  MEL Spectrogram Predictions}.
\newblock \bibinfo{journal}{\emph{2018 IEEE International Conference on
  Acoustics, Speech and Signal Processing (ICASSP)}} (\bibinfo{year}{2018}),
  \bibinfo{pages}{4779--4783}.
\newblock


\bibitem[Shi et~al\mbox{.}(2022)]%
        {av-hubert}
\bibfield{author}{\bibinfo{person}{Bowen Shi}, \bibinfo{person}{Wei-Ning Hsu},
  \bibinfo{person}{Kushal Lakhotia}, {and} \bibinfo{person}{Abdel rahman
  Mohamed}.} \bibinfo{year}{2022}\natexlab{}.
\newblock \showarticletitle{Learning Audio-Visual Speech Representation by
  Masked Multimodal Cluster Prediction}.
\newblock \bibinfo{journal}{\emph{ArXiv}}  \bibinfo{volume}{abs/2201.02184}
  (\bibinfo{year}{2022}).
\newblock


\bibitem[Vaswani et~al\mbox{.}(2017)]%
        {Vaswani2017AttentionIA}
\bibfield{author}{\bibinfo{person}{Ashish Vaswani}, \bibinfo{person}{Noam~M.
  Shazeer}, \bibinfo{person}{Niki Parmar}, \bibinfo{person}{Jakob Uszkoreit},
  \bibinfo{person}{Llion Jones}, \bibinfo{person}{Aidan~N. Gomez},
  \bibinfo{person}{Lukasz Kaiser}, {and} \bibinfo{person}{Illia Polosukhin}.}
  \bibinfo{year}{2017}\natexlab{}.
\newblock \showarticletitle{Attention is All you Need}.
\newblock \bibinfo{journal}{\emph{ArXiv}}  \bibinfo{volume}{abs/1706.03762}
  (\bibinfo{year}{2017}).
\newblock


\bibitem[Vougioukas et~al\mbox{.}(2019)]%
        {vougioukas2019video}
\bibfield{author}{\bibinfo{person}{Konstantinos Vougioukas},
  \bibinfo{person}{Pingchuan Ma}, \bibinfo{person}{Stavros Petridis}, {and}
  \bibinfo{person}{Maja Pantic}.} \bibinfo{year}{2019}\natexlab{}.
\newblock \showarticletitle{Video-Driven Speech Reconstruction using Generative
  Adversarial Networks}.
\newblock \bibinfo{journal}{\emph{arXiv preprint arXiv:1906.06301}}
  (\bibinfo{year}{2019}).
\newblock


\end{thebibliography}










\end{document}